\documentclass[sn-mathphys,Numbered]{sn-jnl}

\usepackage{graphicx}
\usepackage{textcomp}
\usepackage[dvipsnames]{xcolor}
\usepackage{subfig}
\usepackage{multirow}
\usepackage[inline]{enumitem}
\usepackage{relsize}
\usepackage{bbm}
\usepackage{amsmath,amssymb,amsfonts}%
\usepackage{amsthm}%
\usepackage{mathrsfs}%
\usepackage[title]{appendix}%
\usepackage{xcolor}
\usepackage{manyfoot}%
\usepackage{booktabs}%
\usepackage{algorithm}%
\usepackage{algorithmicx}%
\usepackage{algpseudocode}%
\usepackage{listings}
\graphicspath{ {./images/} }

\begin{document}
%
%\title{A Study of Political Leaning Prediction of Social Media Posts Using News Media Bias}
\title[Modeling Political Orientation]{Modeling Political Orientation of Social Media Posts: An Extended Analysis\footnote{This paper is partly funded by the NSF Research Experience for Undergraduates (REU) grant (\emph{Award No.: 2050978})}}

\author*[1]{\fnm{Sadia} \sur{Kamal}}\email{sadia.kamal@okstate.edu}

\author[1]{\fnm{Brenner} \sur{Little}}\email{brenner.little@okstate.edu}

\author[1]{\fnm{Jade} \sur{Gullic}}\email{jade.gullic@okstate.edu}
\equalcont{These authors contributed equally to this work.}

\author[2]{\fnm{Trevor} \sur{Harms}}\email{harmsta@student.swosu.edu}
\equalcont{These authors contributed equally to this work.}

\author[3]{\fnm{Kristin} \sur{Olofsson}}\email{Kristin.Olofsson@colostate.edu}

\author[1]{\fnm{Arunkumar} \sur{Bagavathi}}\email{abagava@okstate.edu}

\affil[1]{\orgdiv{Department of Computer Science}, \orgname{Oklahoma State University}, \orgaddress{\city{Stillwater}, \state{Oklahoma}}}

\affil[2]{\orgdiv{Department of Computer Science}, \orgname{Southwestern Oklahoma State University}, \orgaddress{\city{Weatherford},\state{Oklahoma}}}

\affil[3]{\orgdiv{Department of Political Science}, \orgname{Colorado State University}, \orgaddress{\city{Fort Collins}, \state{Colorado}}}
%
%\titlerunning{Abbreviated paper title}
% If the paper title is too long for the running head, you can set
% an abbreviated paper title here
%

%\author[1]{Sadia Kamal}\email{sadia.kamal@okstate.edu}

%\affil[1]{\orgdiv{Department of Computer Science}, \orgname{Oklahoma State University}, \orgaddress{\city{Stillwater}, \state{Oklahoma}}}

% \author{Sadia Kamal[1] \and
% Brenner Little \and
% Trevor Harms \and
% Jade Gullic \and
% Kristin Olofsson \and
% Arunkumar Bagavathi}
% %

% % First names are abbreviated in the running head.
% % If there are more than two authors, 'et al.' is used.
% %
% \institute{Oklahoma State University, Stillwater, USA \\
% \email{\{sadia.kamal,jade.gullic,abagava\}@okstate.edu}}
% %

% \newcommand{\arun}[1]{\textcolor{red}{\underline{\textbf{\textit{Arun writes:}}} #1}}

%\EarlyAcknow{This project is partly funded by NSF grants.}

\abstract{Developing machine learning models to characterize political polarization on online social media presents significant challenges. These challenges mainly stem from various factors such as the lack of annotated data, the presence of noise in social media datasets, and the sheer volume of data. The common research practice typically examines the biased structure of online user communities for a given topic or qualitatively measuring the impacts of polarized topics on social media. However, there is limited work focusing on analyzing polarization at the ground-level, specifically in the social media posts themselves. Such existing analysis heavily relies on annotated data, which often requires laborious human labeling, offers labels only to specific problems, and lacks the ability to determine the near-future bias state of a social media conversations. Understanding the degree of political orientation conveyed in social media posts is crucial for quantifying the bias of online user communities and investigating the spread of polarized content. 

In this work, we first introduce two heuristic methods that leverage on news media bias and post content to label social media posts. Next, we compare the efficacy and quality of heuristically labeled dataset with a randomly sampled human-annotated dataset. Additionally, we demonstrate that current machine learning models can exhibit improved performance in predicting political orientation of social media posts, employing both traditional supervised learning and few-shot learning setups. We conduct experiments using the proposed heuristic methods and machine learning approaches to predict the political orientation of posts collected from two social media forums with diverse political ideologies: \emph{Gab} and \emph{Twitter}.}


\maketitle              % typeset the header of the contribution
%

%The abstract should briefly summarize the contents of the paper in 15--250 words.

%\input{Abstract.tex}

\section{Introduction}
Online news media forums and social networks play an important role as information system for rapidly disseminating current news to a global audience. However, it is important to recognize that these platforms can significantly impact individuals' political opinions based on the type and quantity of news they present. With millions of people actively engaged in social media forums, the influence and reach of these platforms have become substantial. 
Several factors contribute to the shaping of political ideology among social media users, ranging from the spread of disinformation~\cite{sikder2020minimalistic,10.1145/3578503.3583597}, media bias of news media houses~\cite{bernhardt2008political}, to cognitive bias observed in social media users~\cite{nair2021diffusion}. These factors collectively contribute to polarization, encompassing a spectrum from the \emph{far-left} to the \emph{far-right}~\cite{vicario2019polarization}, among social media users. Exposure to such polarization not only develops online disagreements and ideology segregation but also  has the potential to trigger offline extreme or even violent activities. Therefore, it is essential to develop quantitative methods, leveraging advancements in data science and machine learning, to effectively characterize social media polarization. 

The existing computational approaches study online political polarization in multiple aspects using machine learning with both supervised~\cite{nair2021diffusion} and unsupervised~\cite{fagni2022fine} approaches. However, the existing literature focuses on analyzing political polarization at the user-level, topic-level or news-level. Characterizing the political leaning of social media posts has the potential to become a pre-cursor for hate speech detection, fake news detection ~\cite{aimeur2023fake}, and influence prediction on social media~\cite{vicario2019polarization}. In this work, we present methods to identify political leaning ($\mathcal{P} \in \{\text{left, center, right}\}$) at a fine-grained social media posts using supervised machine learning. Although the existing works use various names like political leaning, political bias, and political orientation, we consider all these terms to be equivalent. In this paper, we use these terms alternatively. Since there are no existing labeled data for supervised political polarization prediction at the posts-level, we present heuristics to assign labels for social media posts in this paper. News media houses take a political stand with their interpretation of topics in news articles. We utilize the political leaning of such news media houses as surrogate data to heuristically label social media posts. 
For example, they identify polarized user communities and topics on social media~\cite{bagavathi2019examining}, with both supervised and unsupervised approaches. 

We experiment with multiple text representation learning frameworks and supervised machine learning models for the political leaning prediction task.  We use existing social media posts from Twitter~\cite{brena2019news} and Gab~\cite{fair2019shouting} and compare the performance of the proposed methods between two datasets. In this work, our contributions are four-fold:

\begin{enumerate}
	\item We present two heuristic methods that are based on news media bias and post sentiment to identify the political leaning of social media posts on Twitter and Gab platforms
    \item We evaluate the heuristically labeled data with the hand-curated data which are labeled by the domain expert and show that the proposed heuristic approaches match significantly with the expert data
	\item We measure the performance of traditional machine learning algorithms to predict the political orientation of posts using a diverse set of text representation learning methods. We compare the performance of machine learning classifiers with different representation learning methods and also between two datasets
    \item We analyze the effectiveness of few-shot learning methods in predicting the political orientation of social media posts with only a very few training data

\end{enumerate}

\section{Related Work}

Social media produces large amounts of data that can be exploited to extract valuable information containing human interactions and opinions. Recent years have seen growing concern that social media forums such as Twitter and Gab may cause political bias in people, affecting presidential elections~\cite{9026882} and news consumption~\cite{garimella2021political}. The detection of political ideology has been a persistent and significant problem, posing considerable challenges. In this section, we discuss several methodologies to identify political leaning. There has been many works that analyzes the problem of political leaning detection in different granularities like sentence-level~\cite{lei2022sentence}, news article level ~\cite{sinno2022political}. Over the years, political leaning detection on social media posts has indeed received less attention. However, there are still some studies and efforts to explore and advance the field of political leaning detection on social media platforms. Textual information is heavily utilized in certain studies ~\cite{6113114,10.5555/3172077.3172399} to identify the political inclination present in social media platforms. But most of these work only utilizes information associated with user behavior ~\cite{10.1145/3503161.3547898}.  
This study~\cite{preotiuc-pietro-etal-2017-beyond} unveil the variations in language patterns exhibited by different ideological groups and develops an algorithm for predicting the political ideology of individual users, considering all levels of engagement. Retweet-BERT ~\cite{jiang2023retweet} captures a comprehensive understanding of user ideology by combining network structure and language cues. It involves training in an unsupervised manner on the complete dataset, they also learn the representations based on user's profile descriptions and retweet interactions. This study~\cite{hosseinmardi2021examining} also analyzes political polarization by mapping user information to learn user behavior. While many existing works focus on utilizing user details to detect political bias, our approach differs as we specifically leverage the content of the post itself for the political orientation identification. Instead of relying on user information, we employ two heuristic methods to label the posts based on their content.

Machine learning approaches have become increasingly popular for detecting political bias using traditional lexicon-based classifiers based on "bag-of-words" techniques ~\cite{10.5555/3104482.3104544}. There are several problems with these approaches, including the overreliance on primary-level lexical information and the neglect of semantic structure. Another study~\cite{olorunnimbe2015tweets} utilized machine learning models such as Naive Bayes algorithm, to classify users based on their political inclinations. Sentiment or opinion-based techniques leverage natural language processing (NLP) or text mining approaches to comprehend people's opinions towards political figures. These techniques involve analyzing the textual content of posts to calculate a score indicating the sentiment expressed. Some researchers used NLP-based methods ~\cite{chen-etal-2018-learning}, and 
attention-based multi-view model ~\cite{kulkarni-etal-2018-multi} to identify the political leanings of topics. This study ~\cite{xiao2023detecting} focuses on examining political bias in text corpora by curating a dataset sourced from Twitter and introduce PEM model, that captures and analyzes both the semantic meaning and political polarity conveyed within the text. The work in ~\cite{pandya2022proposal}studies political bias detection in news media articles using state-of-the-art NLP methods like BERT and analyzing the text similarity.  This paper also  ~\cite{glazkova2021comparison} utilizes text representation methods for identifying political viewpoint of social media users based on textual data. To compare different document representation choices,  deep learning approaches,  semantically meaningful word embeddings and attention mechanisms are analyzed in ~\cite{cruz2020document}.This work~\cite{osti_10178656} presents a methodology to identify political ideology on the Twitter platform by capturing and analyzing the interactions between various relationships within the network and assessing the significance of each relationship in the context of ideology detection. Another recent work employed a methodology IOM-NN (Iterative Opinion Mining using Neural Networks)~\cite{9026882} by leveraging neural networks to estimate the polarization of public opinion regarding political events. Some of the works utilized statistical measures for news media bias~\cite{sales2019media}, fair models for political leaning prediction on news articles~\cite{baly-etal-2020-detect,baly-etal-2019-multi}, and examining an online community behavior with multi-modal data~\cite{hosseinmardi2021examining}  This paper ~\cite{gangula-etal-2019-detecting} detects political bias in news media headlines using attention mechanism and NLP techniques. This paper~\cite{ren2022discrimination} introduces a graph neural model along with a heterogeneous network to model news content for evaluating the political leaning of news. Hierarchical attention networks are utilized by this paper~\cite{10.1145/3543507.3583300}  the to employ KHAN (Knowledge-based Hierarchical Attention Networks) for political stance prediction in a news article. Other than attention based models, neural network models some work~\cite{chakraborty2022fast} identifies political inclination using unsupervised learning methods like fast few-shot method. In this study, we propose two heuristic methods for identifying political bias at the post-level. Additionally, we investigate the effectiveness of machine learning and few-shot learning methods for the identification task.

\section{Datasets}
In this work, we use web resources to capture the political leaning of online news domains along with existing datasets collected from social media forums. We describe the datasets used in our experiments and methods in this section. We publish all the datasets used in this paper at our Github repository\footnote{\url{https://github.com/sadiakamal/Tweets_Political_Orientation/tree/main}} for the research community to explore future directions of this research.

\begin{figure}[h]
   \centering
    \includegraphics[scale=0.7]{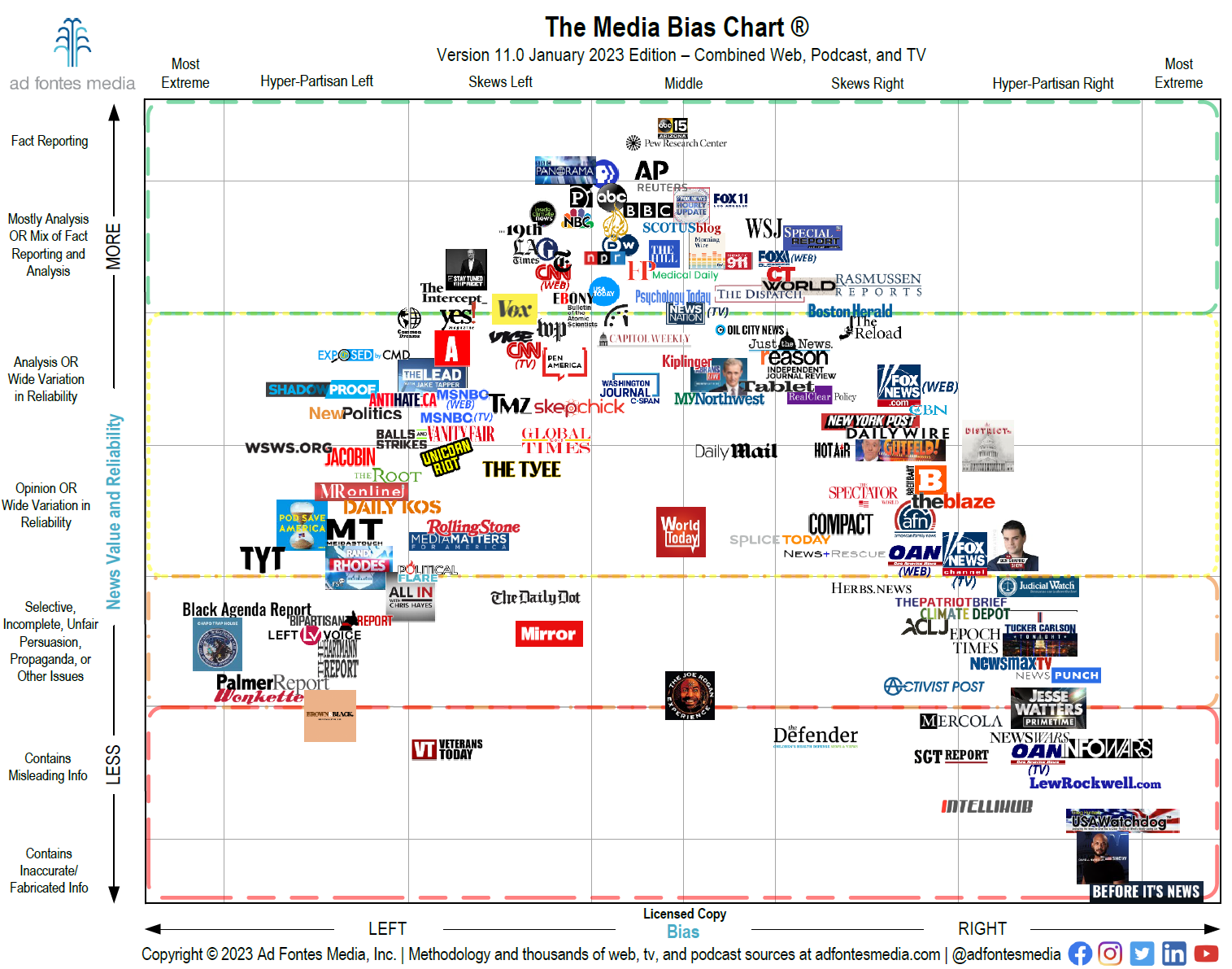}
    \caption{Media bias chart of allsides.com. This chart illustrates the number of news domains in their repository along with their political orientation and reliability}
    \label{fig:Bias_Chart}
\end{figure}

\subsection{News Domains Data}

We leverage the utilization of political orientation present in news domains to associate the political leaning of social media posts. Thus first, we collect the political leaning ($\mathcal{P} \in \{\text{left, center, right}\}$) of news domains. Given a set of \emph{n} news domains $\mathcal{D} \in \{d_1,d_2,\ldots,d_n\}$, we associate a political leaning of each news domain $d_i \in \mathcal{D}$ with one of the political leaning labels using $d_i \rightarrow \mathbbm{1}_\mathcal{P}$. With the help of web scrapping tools, we created a media bias dataset from allsides.com \footnote{\url{http://www.allsides.com/media- bias/ media-bias-rating-methods}}. \emph{allsides.com} provides a diverse viewpoint on news articles and analyzes the political leaning of all online news media domains with crowd sourcing. Figure~\ref{fig:Bias_Chart} depicts the distribution of several news domains that discuss US politics like \emph{CNN, Fox News, CNBC}, and \emph{nytimes} available in the repository with their political orientation given in the scale from \emph{far-left} to \emph{far-right}. Our media bias dataset contains a total of 422 news domains, out of which \emph{158} news domains are labeled \emph{left}, \emph{166} are labeled \emph{center}, and \emph{98} are labeled \emph{right}. It is important to note that we merged news domains with political orientation of \emph{far-left} and \emph{far-right} to \emph{left} and \emph{right} categories respectively. This is to mitigate the uneven data distribution problem with \emph{far-left} and \emph{far-right} labels. 
%The dataset collected from \emph{allsides.com} contains online news domains like \emph{CNN, Fox News, CNBC}, and \emph{nytimes} along with their corresponding political leaning. 

\subsection{Social media posts} In our work, we use publicly available datasets which are from social media sites Twitter~\cite{brena2019news} and Gab~\cite{fair2019shouting} for all experiments. 

\begin{figure}[h]
    \centering
    \subfloat[\centering News Domains in Gab \label{fig:NewsGab}]{{\includegraphics[scale=0.4]{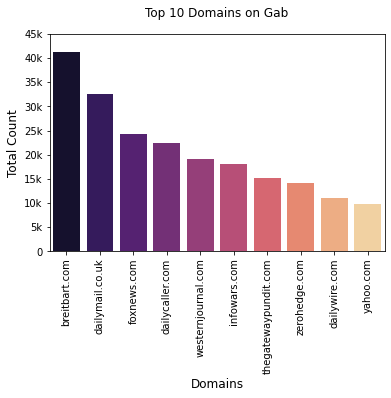} }}
    \qquad
    \subfloat[\centering  News Domains in Twitter \label{fig:NewsTweet}]{{\includegraphics[scale=0.4]{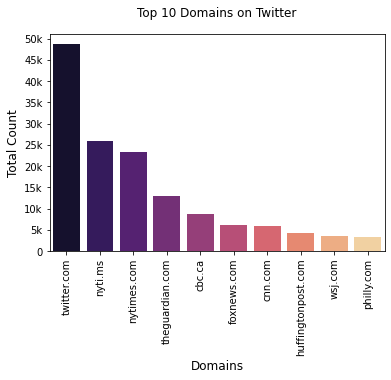} }}
    \caption{Top 10 news domains in Gab and Twitter datasets}
    \label{fig: Top 10 news domains}
\end{figure}

\subsubsection{Twitter} Tweets with news article URLs that discuss political topics from mainstream news media sources are included in the Twitter dataset~\cite{brena2019news}. The timeline of the Twitter dataset is from January 2018 to September 2018. %This dataset is constructed using Newspaper3k library and tweepy to enable Twitter search API. 
The dataset comprises \emph{289,738} tweets with URLs from \emph{60} handpicked mainstream news domains covering a diverse range of political views. 
%This dataset is constructed by collecting tweets that contain 69 manually picked popular U.S. online news media houses.%which include 32 newspapers and 37 news agencies. 
%Most recent tweets that contain the news URLs explicitly are collected.
%The dataset consists of news articles collected from multiple news domains covering a diverse range of political views and public tweets referring to such articles.
%The dataset comprises news articles collected from news domains that range from a wide spectrum of political leaning and public tweets that mention such news articles. 

\subsubsection{Gab} The Gab dataset~\cite{fair2019shouting} contains a  total of 40 million posts including posts, replies, and re-posts collected in 2018 which mostly support far-right ideologies. We further select only the posts with more than five words along with news URLs to get post context on news articles. In total, our dataset contains \emph{1,368,028} posts with URLs from \emph{355} news domains.
% The social media platform Gab typically supports free-speech and far-right ideologies~\cite{bagavathi2019examining}.
% We only use Gab posts from January 2018 to September 2018, to match the Twitter dataset timeline.

\subsubsection{Ground Truth Data}
We are building labeled datasets required for our experiments using heuristic methods as detailed in Section~\ref{sect:methodology}. We construct a small ground truth data which contains a small sample of social media posts labeled by a domain expert to evaluate our heuristic approaches. To construct this data, we randomly select \emph{161} posts with news article URL(s) from our Twitter dataset. Next, these samples are given to a political science expert, who carefully analyzed the content of each tweet and assign a label. This assignment is based on the post's alignment or disagreement with the news article mentioned in the post. Specifically, if the content of the tweet is in agreement with a particular media outlet's political leaning, it is labeled as having that news domain's political orientation. Conversely, if the tweet's content  is in disagreement with a media outlet's leaning, it is labeled as having the opposite bias. It is important to note that we label the post as \emph{Center}, even if the post is in disagreement with \emph{Center} leaning news domain due to uncertaininty. This is one of the problems to explore in the future as it requires a comprehensive analysis on news articles along with news domains' political orientation. This expert annotation serves as a benchmark to evaluate the accuracy of our labeling methods.

%To assess the accuracy of our labeling methods compared to expert annotation, we conducted a user study.

Figures ~\ref{fig:NewsGab} and ~\ref{fig:NewsTweet} show the top ten news domains appearing in both Twitter and Gab datasets. We can see that the \emph{Gab} dataset contains large number of news articles from right-aligned news media outlets like \emph{breitbert}, \emph{dailymail}, and \emph{foxnews}. On the other hand, \emph{Twitter} posts are leaning more towards left aligned mainstream news media forums such as \emph{nytimes}, and \emph{theguardian}.

\section{Methodology}
\label{sect:methodology}

\begin{figure}[htbp]
   \centering
    \includegraphics[scale=0.68]{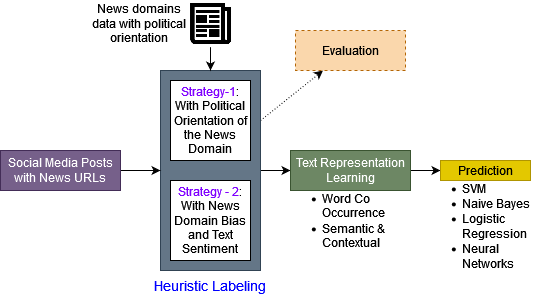}
    \caption{Pipeline of the methodology of this work. The given social media posts are first pre-processed and assigned political orientation label. We introduce two different strategies to heuristically label social media posts. We use both frequency-based and contextual text representation learning methods to extract feature vectors of social media posts. We predict the political orientation of social media posts with traditional machine learning classifiers using the extracted features. We also perform a small scale evaluation for the heuristically labeled data.}
    \label{fig:Pipeline}
\end{figure}

To quantify the characteristics of political polarization on online social media with machine learning approaches, we require annotations of social media posts. Due to the near impossibility of hand curating millions of posts, the existing approaches analyze polarization at the user-level~\cite{hosseinmardi2021examining} and news-level~\cite{baly-etal-2020-detect}. The methodology used in this paper is summarized in Figure~\ref{fig:Pipeline}. We give description of heuristic labeling in this section and we outline the details of text representation learning and classification methods used in our experiments in Section~\ref{sect:results}.

\subsection{Heuristic Data Labeling}
In this paper, we propose two heuristic methods to label the political leaning of social media posts on Twitter and Gab. Our proposed heuristic approach does not require any human-annotated data and we base these heuristics on the available news media bias dataset collected from allsides.com.

\subsubsection{Type-1: News Domain Labeling}

Our first heuristic approach to label political leaning of social media posts is based on news media houses and their political leaning. We use our media bias dataset created from allslides.com that has the news media source and their political leaning: \emph{Left} (-1), \emph{Center} (0), and \emph{Right}(1). Given a set of \emph{m} social media posts $\mathcal{S}\in \{s_1,s_2,\ldots,s_m\}$, we give a heuristic to associate each social media post $s_i \in \mathcal{S}$ with one of the political leaning labels: $s_i \rightarrow \mathbbm{1}_\mathcal{P}$. Given a social media post $s_i$ containing one or more news articles from a set of $q$ news domains $\mathcal{\widehat{D}} = \{\hat{d_1},\hat{d_2},\ldots,\hat{d_q}\}$ where $\mathcal{\widehat{D}} \subset \mathcal{D}$ and $q <<< n$, our proposed heuristic approach obtains the political leaning of the post ($\mathcal{P}_{s_i}$) using Equation~\ref{eqn:type-1}.

%$\mathcal{P}_{s_i} = \frac{\[ \sum_{j=1}^{m} \mathcal{P}_{d_j} \]}{m}$
% \mathcal{\widehat{P}}_{s_i} \: ={}& \frac{\sum_{j=1}^{m} \mathcal{P}_{d_j}}{m} \\

\begin{equation}
\begin{split}
	\mathcal{P}_{s_i} \: = \begin{cases}
	-1 \: \: if \; \mathcal{\widehat{P}}_{s_i} < 0.1 \\
	\: \: 0 \: \: \: if \; -0.1 < \mathcal{\widehat{P}}_{s_i} < 0.1 \\
	+1 \: \: if \; \mathcal{\widehat{P}}_{s_i} > 0 \\
	\end{cases}
	\label{eqn:type-1}
\end{split}
\end{equation}

 where $\mathcal{\widehat{P}}_{s_i} = \frac{\sum_{j=1}^{m} \mathcal{P}_{d_j}}{m} \: ; \: -\infty < \mathcal{\widehat{P}}_{s_i} < +\infty$ is an unscaled political leaning of the social media post $s_i$ and $\mathcal{P}_{d_j}$ is the political leaning of news domain $d_j \in \widehat{D}$.  Since we use these labels for supervised machine learning algorithms, we use Equation~\ref{eqn:type-1} to assign final political leaning label of the social media post $\mathcal{P}_{s_i}$. We define our heuristics for social media posts with news domain URLs and this method will not assign any labels if there are no news domain URLs in social media posts.
 
%In other words, we assign the average of sum of political leaning ($\mathcal{P}_{d_j}$) of each news domain in $d_j \in \widehat{D}$ as the unscaled political leaning ($\mathcal{\widehat{P}}_{s_i}$) of the social media post $s_i$. 
 
%We iterate over the posts to find news media URLs  $\mathcal{U}$, where $\mathcal{U}\in \{u_1,u_2,\ldots,u_k\}$ is a set of all the URLs,  with a heuristic approach we extract the domain $\mathcal{D}$ of the URLs.  We match that domain from the post with the domains present in our Media Bias dataset such that, $\mathcal{P}_{s_i} = \mathcal{P}_{d_j} ; d_j \in s_i$.  If the matching is successful we label the post with the same label as the news domain.

\subsubsection{Type-2: Sentiment Labeling}
Our second heuristic approach to labelling political leaning of social media posts is based on both news media bias and social media post content. The proposed method is an extension of our first approach and it intends to improve the accuracy of the labels by including the post's sentiment in news articles. We measure the sentiment score of social media posts by combining scores collected from three sentiment analyzer tools.

\textbf{Textblob} : Textblob is a popular sentiment analysis tool which returns polarity and subjectivity score. In this paper we use only the polarity score from TextBlob and we represent it as $\alpha \in [-1, +1]$. $\alpha = -1$ refers to negative sentiment and $\alpha = +1$ refers to positive sentiment. Subjectivity score of a sentence refers to the category of the sentence, for example, opinion or emotion etc.

\textbf{Vader}: Vader is another popular sentiment analysis tool which gives the probability of a sentence being positive, negative or neutral. We represent the computed score from Vader as $\beta \in \{-1, 0, +1\}$, where -1 is negative, 0 is neutral and +1 is positive. Vader analyzes the semantic orientation of a sentence and lexical features to calculate the sentiment of the given text.

\textbf{Afinn}: Afinn is constructed with a large corpus of labeled lexicons where each lexicon has a polarity score associated with it. Similar to TextBlob, we represent the polarity score of Afinn as $\gamma \in [-1, +1]$. We use this to get the polarity scores of the words of our social media posts.

Given a set of social media posts $\mathcal{S}$, we compute the sentiment score ($\tau$) of the post $s_i \in \mathcal{S}$ by combining sentiment values from three sentiment analyzer tools as $\tau_{s_i} = \frac{\alpha+\beta+\gamma}{3}$. With a sentiment value of a post $\tau_{s_i}$, we compute the unscaled political leaning ($\mathcal{\widehat{P}}$) as $\mathcal{\widehat{P}}_{s_i} =  \mathcal{\widehat{P}}_{s_i} \times \tau_{s_i}$. We further update the political leaning of the given post using Equation~\ref{eqn:type-1}. In other words, we switch the $\mathcal{P}_{s_i}$ only if the overall sentiment of the post ($\tau_{s_i}$) is negative.
For example if the label is originally Left then we will switch the label to Right. If the label is Center,  there is no change to the label.

\subsection{Machine Learning Training Strategies}
We follow two types of machine learning classification approaches to predict the political orientation of social media posts as described bellow.

\subsubsection{Supervised Learning}
Most of our experiments are conducted in a supervised learning setting. We utilize the heuristically labeled datasets, which we split into an 80-20 ratio for training and evaluation purposes. This approach allowed us to leverage the available labeled data and train our models using conventional supervised learning techniques.

\subsubsection{Few-shot Learning}
In our study, we employed few-shot learning with BART to tackle the issue of limited labeled data in the political leaning detection task.  Few-shot learning~\cite{snell2017prototypical} combines elements of both supervised and unsupervised learning. By leveraging the capabilities of BART, which can effectively learn from a small number of labeled examples, we aimed to develop a more robust representation of political leaning. To train our model, we utilized a small training set consisting of only 1000 samples per class. This limited sample size allowed us to simulate a scenario with scarce labeled data, where obtaining extensive annotations is challenging. The overall dataset we worked with is comprised of a total of 75,000 samples.

\section{Experiments and Results}
\label{sect:results}

In this section, we present the results of our labeling heuristics. Also, we experiment with existing machine learning approaches to predict the political leaning of social media posts. 

\subsection{Political Orientation Labeling}

\begin{figure}[ht]
    \centering
    \subfloat[\centering News Domain Labeling \label{fig:NewsLabel}]{{\includegraphics[scale=0.37]{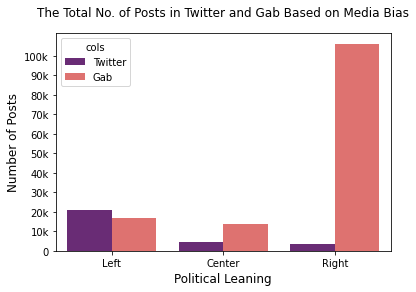} }}
    \qquad
    \subfloat[\centering Sentiment analysis labeling \label{fig:SentimentLabel}]{{\includegraphics[scale=0.37]{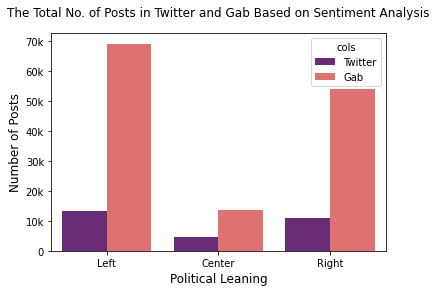} }}
    \caption{Political bias labeling summary with (a) Heuristics-1: based on political bias of news domains and (b) Heuristics-2: based on political bias of news domains and post sentiment}
    \label{fig:Sentiment_analysis_labeling}
\end{figure}

In figure ~\ref{fig:NewsLabel} and ~\ref{fig:SentimentLabel} we show the political leaning labeling summary for both the datasets with our Heuristic 1 and Heuristic 2 labeling methods. Labels collected using our heuristic 1 labeling method correlate with our qualitative analysis given in Figure~\ref{fig: Top 10 news domains}. As depicted in figure~\ref{fig:NewsLabel}, it is evident that Gab posts tend to focus on right-leaning news outlets whereas Twitter posts primarily share articles from left-leaning news sources. However, this scenario changes when we also consider post content using our heuristic 2 as shown in figure ~\ref{fig:SentimentLabel}.  More than $50\%$ of Gab posts support left and center-aligned posts, while most of the posts on Twitter still support left-leaning news.

\subsection{Methods for learning text representations}
\label{sect:rep}

We experiment with five different text representation learning methods to extract features of our labeled social media posts. We choose two text representation learning methods that are based on word co occurrence and three text representation learning methods that extract semantic and contextual features of the text data. Following are the feature extraction methods we used our experiments.
 
\textbf{Bag-of-Words}: Count vectorizer or Bag-of-Words~\cite{shahmirzadi2019text}  model generates vectors of social media posts based on only the occurrence frequency of each word.The feature dimension of each social media post will be equal to the size of the vocabulary.

\textbf{TFIDF Vectorizer}: TFIDF model~\cite{ramos2003using} measures the importance of a word based on the occurrence frequency of words in all social media posts along with the relevance of words to a social media post. The feature length of each social media post from the TFIDF vectorizer will also be equal to the size of the vocabulary.

\textbf{Word2vec}: %The Word2Vec~\cite{DBLP:journals/corr/abs-1301-3781} is a popular model to generate contextual embeddings using word semantics in a sentence. In particular,.
We use the skip-gram version of Word2Vec~\cite{DBLP:journals/corr/abs-1301-3781} model that tries to predict the probability of occurrence of context words, given a keyword $w$ from $T$ words. The context is assigned using a sliding window of size \emph{c}. 
Word2vec aims to keep the contextually similar words close in the feature space compared to dissimilar words.
%using the objective function given in Equation ~\ref{eqn:wordvec}

%\begin{equation}
%\frac{1}{T} \sum_{t=1}^{T} \sum_{-c\le j \le c, j \neq 0} \log p (w_{t+j}|w_t)
%\label{eqn:wordvec}
%\end{equation}
  
\textbf{BERT}: To comprehend the political leaning associated with each post, we employ BERT (Bidirectional Encoder Representations from Transformers)~\cite{devlin-etal-2019-bert} model and conduct a customized training process. The BERT model is pre-trained on a large corpus of text data, allowing it to learn contextual patterns and linguistic nuances present in the language. Importantly the BERT model is pre-trained specifically for two main tasks: masked language model (MLM) and next sentence prediction (NSP). However, we fine-tune the BERT model using our heuristically labeled datasets to make it specifically tailored for our downstream task of political orientation prediction. By fine-tuning the pre-trained BERT model on our dataset, we enhance its ability to capture the connections of political orientation to the context of social media posts.
%By leveraging BERT's advanced language understanding capabilities, we can analyze and interpret the political implications of the posts more effectively

\textbf{BART}: BART (Bidirectional and AutoRegressive Transformers) ~\cite{lewis-etal-2020-bart} is a state-of-the-art language generation model that is based on the Transformers architecture. BART is primarily focused on understanding language, BART is designed for generating coherent and contextually relevant text. Although text classification is not one of the primary purposes of the BART model, it can be fine-tuned the same way as BERT for the political orientation prediction task. We employ the BART model only for the few-shot learning strategy.

\subsection{Political Leaning Prediction}

We use traditional machine learning models to predict political leaning $\mathcal{P}$ of social media posts using text features extracted by one of text representation learning methods given in Section~\ref{sect:rep}. We give details about classifier models used in our experiments and their hyperparameters below:

\textbf{SVM}: The goal of a support vector machine(SVM) is to find a hyperplane in an N-dimensional space to classify the nearest data points. The goal of this classifier is to maximize the distance between data points and the hyperplane. We use SVC class from sklearn library with \emph{C=1.0} and \emph{RBF} kernel.

\textbf{Logistic regression}: In a logistic regression classifier, the relationship between the independent and the dependent variables is estimated by applying a predictive modeling approach. We use logistic regression module from sklearn package with the default hypermeters \emph{C=1.0} and solver as \emph{lbfgs}.
%In a logistic regression classifier, the relationship between the independent and the dependent variables is estimated by applying a predictive modeling approach. 

\textbf{Naive Bayes}: We particularly use \emph{GaussianNB} in our experiments. 
%We use naive bayes classifier for the classification task and it uses the Bayes theorem also they are easy in terms of implementation. %but as the predictors are dependent, it deters the performance of the classifier.

\textbf{Neural network}: We use a sequential neural network architecture with one input layer, three Dense layers and one classifier head which is fine-tuned with \emph{Categorical Crossentropy} loss function and \emph{Adam} optimizer and a learning rate of 1e-06. As the semantic text representation learning models like BERT and BART are based on neural network architecture, we use only the neural network classifier for these models.

\begin{table}[h!]
\centering
\caption{Performance of traditional classifier models in accuracy ($\%$) with News Domain labeling based on text features extracted using word frequency, TFIDF, and Word2Vec}
\begin{tabular} { p {3cm}|p {1.5cm}|p {1.5cm}|p {1.5 cm}|p {1.5 cm}}
\hline
\multicolumn{5} { c  }{Twitter Data}\\
\hline
---- & SVM &LR & Naive Bayes & NN \\
\hline
W2V & 0.44 & 0.50 & 0.40 & 0.55\\

%FScore & 0.39 & 0.40 & 0.36 & \textbf{0.51}\\

%---- & SVM &LR & Naive Bayes & NN \\
\hline
Count & 0.628 & 0.663 & 0.508 & 0.64 \\%\textbf{0.70}\\

%FScore & 0.60 & 0.60 & 0.56 & \textbf{0.68}\\

\hline

TFIDF & 0.671 & 0.675 & 0.524 & 0.635 \\%\textbf{0.71}\\
%FScore & 0.60 & 0.60 & 0.57 & \textbf{0.68}\\
%\hline
%BERT & 0.63 & 0.56 & 0.37 & 0.61\\

%FScore & 0.56 & 0.50 & 0.37 & \textbf{0.54}\\
\hline
\hline
\multicolumn{5} {  c  }{Gab Data}\\
\hline
W2V  &  0.51 & 0.50 & 0.44 & 0.62\\
\hline
Count &  0.915 & 0.949 & 0.585 & 0.913 \\%\textbf{0.97}\\
\hline
TFIDF & 0.913 & 0.909 & 0.548 & 0.862\\ %\textbf{0.96}\\
%\hline
%BERT & 0.75 & 0.65 & 0.41 & 0.75\\
\hline

\end{tabular}
\label{table:NewsDomain}
\end{table}

\begin{table}[h!]
\centering
\caption{Performance of traditional classifier models in accuracy ($\%$) with sentiment analysis labeling based on text features extracted using word frequency, TFIDF, and Word2Vec}
\begin{tabular} { p {3 cm}|p {1.5 cm}|p {1.5 cm}|p {1.5 cm}|p {1.5 cm}}
\hline
\multicolumn{5} { c  }{Twitter Data}\\
\hline
---- & SVM &LR & Naive Bayes & NN \\
\hline
W2V & 0.42 & 0.41 & 0.39 & 0.51\\

\hline
Count & 0.527 & 0.564 & 0.486 & 0.521\\

\hline

TFIDF & 0.56 & 0.571 & 0.47 & 0.541\\%\textbf{0.68}\\
\hline
\hline
\hline
\multicolumn{5} { c }{Gab Data}\\
\hline
W2V  &  0.51 & 0.50 & 0.44 & 0.62\\
\hline
Count &  0.72 & 0.734 & 0.483 & 0.683\\%\textbf{0.85}\\
\hline
TFIDF & 0.705 & 0.707 & 0.497 & 0.664\\
\hline

\end{tabular}
\label{table:SentimentLabel}
\end{table}

% In Tables ~\ref{table:NewsDomain} and ~\ref{table:SentimentLabel} we present the accuracy results of machine learning models using each text representation model to predict the political leaning of posts collected from Twitter and Gab. The results in Table ~\ref{table:NewsDomain} show the accuracy of each classifier for the data labeled using our first heuristic approach. In general, we notice that the TFIDF feature set yields the best results for the Twitter dataset, and the BoW feature set yields the best results for the Gab dataset. Based on the results, we can see that Gab has a significantly higher accuracy score than Twitter due to the availability of a large quantity of diverse data. In addition, we observe that of all the models, our neural network model gives the best classification performance because of the hidden layer architecture, which captures the complexity of the data.

It is evident from Tables~\ref{table:NewsDomain} and~\ref{table:SentimentLabel} that neural networks and logistic regression classifier models outperform other machine learning models to predict the political leaning of Twitter and Gab posts. It is interesting to note that the models trained with feature extraction methods such as Bag-of-Words (BoW) and TF-IDF achieve better performance than models trained with semantic text representation learning methods like Word2Vec for the data that are labeled with Heuristics-1. However, we emphasize that the Word2Vec model gives almost equal performance with classifiers when trained with the data labeled using the Heuristics-2 strategy. It is also important to note that all prediction models experience a significant drop in performance (approximately $11\%$ accuracy) for the Gab dataset when using the Heuristics-2 labels. These findings suggest that the labeling method may have a considerable impact on the model's performance, particularly for the Gab dataset.

\begin{table*}[h]
\centering
\caption{Neural Networks classifier performance with BERT representations} 
\label{table:Bert} 

\begin{tabular}{p{3 cm}p{1.5cm}| p{1.5cm} p{1.5cm} p{1.5cm} p{1 cm}}
 \hline
 \hline
\multicolumn{2}{c|}{\multirow{1}{*}{Dataset}} & Acc. (\%) &Macro F1& Precision & Recall \\
\hline
News Domain & Twitter & 0.7613 & 0.7598& 0.7618 & 0.7682\\
&Gab& 0.9942&0.9942&0.99426& 0.99429\\
Sentiment & Twitter & 0.6844 &  0.6848& 0.6854&0.7054 \\
& Gab & 0.8896&  0.8891& 0.88919& 0.8894\\

\hline

\end{tabular}
\end{table*}

Table \ref{table:Bert} demonstrates the performance results of the state-of-the-art text representation model BERT.
The results clearly indicate that BERT outperforms all other text representation models presented in Tables ~\ref{table:NewsDomain} and ~\ref{table:SentimentLabel} for both heuristic strategies and both datasets. This performance can be attributed to BERT's contextualized word embeddings, bidirectional representation learning, and its ability to capture domain-specific context effectively after fine-tuning. Specifically, BERT achieved an accuracy of 76$\%$ for the Twitter dataset and an accuracy of 99$\%$ for the Gab dataset. We also note that the model performance drops when using the Hueristics-2 strategy of sentiment labels.

\begin{table*}[h]
\centering
\caption{Performance metrics of the neural network classifier trained in a few-shot learning setup with BART text representations.} 
\label{table:Bart} 

\begin{tabular}{p{3 cm}p{1.5cm}| p{1.5cm} p{1.5cm} p{1.5cm} p{1 cm}}
 \hline
 \hline
\multicolumn{2}{c|}{\multirow{1}{*}{Dataset}} & Acc. (\%) &Macro F1& Precision & Recall\\
\hline
News Domain & Twitter & 0.72 & 0.61& 0.52 & 0.72\\
&Gab& 0.33&0.17&0.11& 0.33\\
Sentiment & Twitter & 0.46 &  0.29& 0.22&0.46\\
&Gab& 0.33&0.17&0.11& 0.33\\

\hline

\end{tabular}
\end{table*}

Table \ref{table:Bart} presents the results of few-shot learning using the text representation model BART. While BART performed well for the Twitter dataset, it did not achieve satisfactory performance for the gab datasets using both labeling methods.

\begin{figure}[h]
    \centering
    \subfloat[\centering Twitter Dataset \label{fig:NewsGab_mat}]{{\includegraphics[scale=0.29]{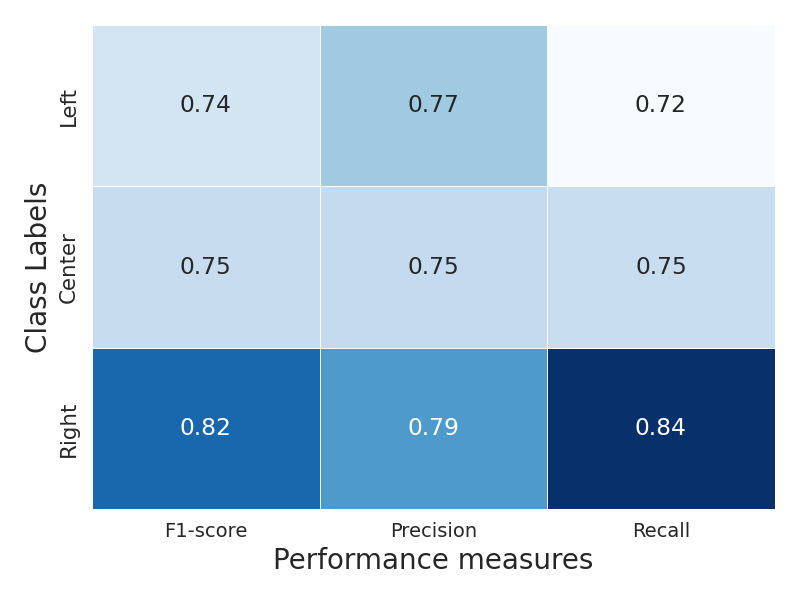} }}
    \qquad
    \subfloat[\centering Gab Dataset\label{fig:NewsTweet_mat}]{{\includegraphics[scale=0.29]{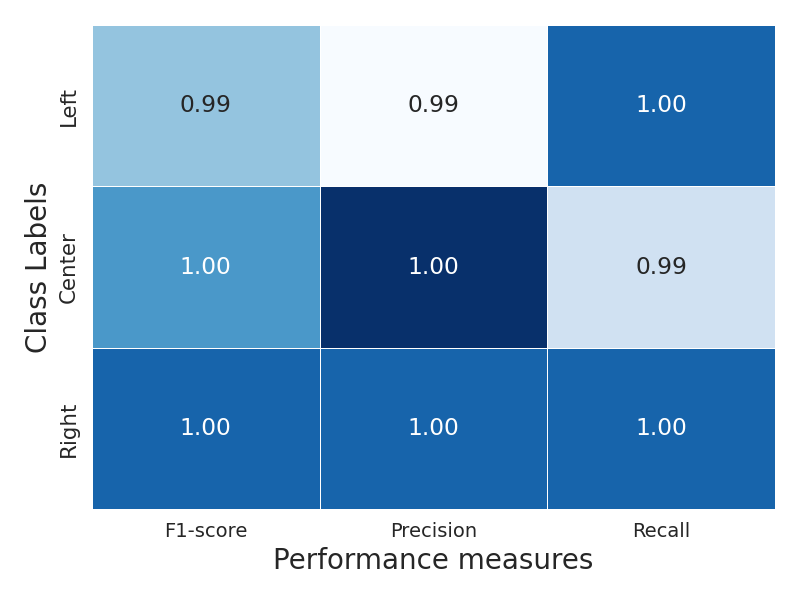} }}
    \caption{The BERT model's performance on Heuristics-1 labeled datasets for each class}
    \label{fig: Newsdomain_mat}
\end{figure}

% Table ~\ref{table:SentimentLabel} shows the results of each classifier's accuracy using our second heuristic approach. Although the accuracy of ML models is relatively low for this datasets, we notice that the ML algorithms can identify political leaning of Gab posts better than that of Twitter posts. 

% . We also note that machine learning models trained with features extracted using simple methods like BoW and TFIDF outperform models trained with state-of-the-art text representation learning methods like Word2Vec. While we note that models on Gab perform better the accuracy of all models decreases by about $14\%$ for the Gab dataset when we use the Heuristics-2 labels.

\begin{figure}[h]
    \centering
    \subfloat[\centering  Twitter Datset\label{fig:SentTweet_mat}]{{\includegraphics[scale=0.29]{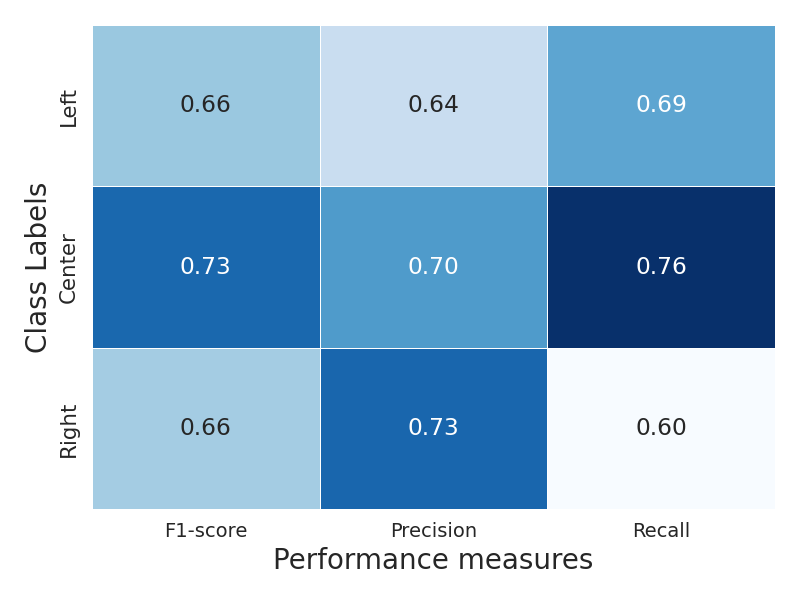} }}
    \qquad
    \subfloat[\centering Gab Datset \label{fig:SentGab_mat}]{{\includegraphics[scale=0.29]{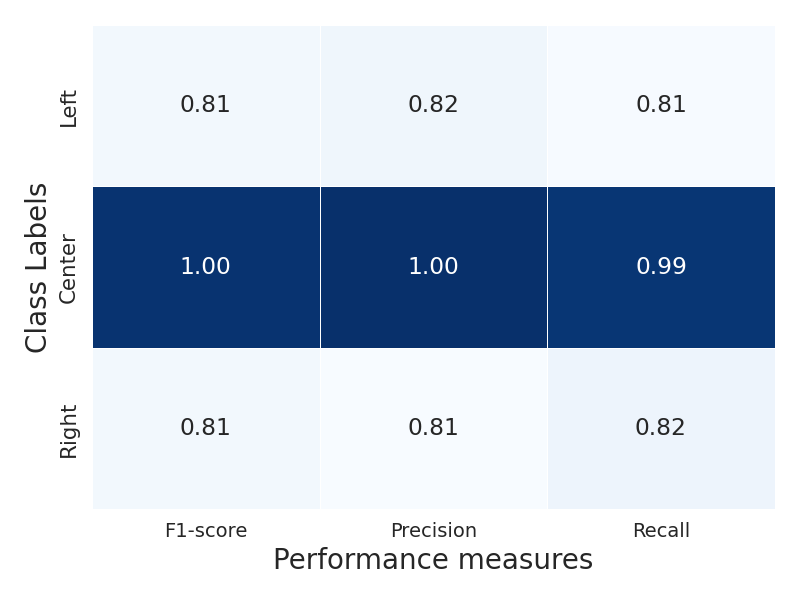} }}
    \caption{The BERT model's performance on Heuristics-2 labeled datasets for each class}
    \label{fig: Sentiment_mat}
\end{figure}

In Figures \ref{fig: Newsdomain_mat} and \ref{fig: Sentiment_mat} we report the performance measures(f1-score, precision, recall) for all the target classes \emph{(left, center and right)}. As clearly illustrated in Figures \ref{fig: Newsdomain_mat} and \ref{fig: Sentiment_mat} the model correctly identified most of the posts belonging to the \emph{right} class for all the datasets except Heuristics-2 labeled gab dataset. Each heatmap tends to be darker for the \emph{right} labels.

\begin{table*}[h]
\centering
\caption{Evaluation of heuristically labeled social media posts using hand-curated data. The numbers in each column represents total number of heuristic labels that correctly correlates with hand-curated labels. Note that there is 100\% correlation with the \emph{Center} label as the current hand-curated data does not have resources to consider disagreements with \emph{Center} news domains.} 
\label{table:GT} 

\begin{tabular}{p{3 cm}p{1.5cm}| p{1.5cm} p{1.5cm} p{1.5cm}|p{1cm}}
 \hline
 \hline
\multicolumn{2}{c|}{\multirow{1}{*}{Dataset}} & Left & Center& Right & Total\\
\hline
News Domain & Twitter & 49 & 41& 61 & 151\\
Sentiment & Twitter& 25 & 41& 17 & 83\\

\hline

\end{tabular}
\end{table*}

Table \ref{table:GT} presents an evaluation for the heuristically labeled datasets using the expert-labeled Twitter data. We do not possess the hand curated datasets for the Gab posts due to limited resources and time. We will set this task for future experiments. We note from Table \ref{table:GT} that Heuristics-1 labeling method shows a 93.78 $\%$  match, with 151 out of 161 samples aligning with hand-curated data. However, Heuristics-2 labeling method shows only a 51.55 $\%$ match with 83 out of 161 samples aligning. The results indicate that users sharing articles mostly align their political leaning with the news domain's political leaning. This also supports our Heuristics-1 labelling strategy for approximating obtaining political orientation labels of Twitter posts.

\section{Conclusion and Future work}
In this paper, we provided two heuristic methods to approximate political orientation in social media posts. We obtained political leanings of social media posts from the large-scale Twitter and Gab datasets using a surrogate news media bias dataset. We also evaluated a sample of these heuristic labels by comparing them with hand-curated tweets and analyzed that one of our heuristic methods give better approximations in identifying the political orientation. Finally, we explore state-of-the-art text representation techniques to extract the features to train traditional machine learning models to predict the political leaning of any given post. This study opens several avenues to exploit in the future:

\begin{enumerate}
    \item \textbf{Direction - 1:} Although we added an evaluation with a small hand-curated dataset, this study require a large scale analysis. The future contributions could add more effort in collecting more hand-curated data for a robust evaluation on the proposed heuristic methods. Another future contribution can be hand labeling with in-depth analysis of both social post and the attached news article to decide the final political orientation.
    \item \textbf{Direction - 2:} The current method is only designed for handling the text data. The proposed heuristics fail if there are no text in the social media post. However, social media posts come in multiple modalities like text, networks, and images. Developing methods that suit multi-modal data can improve label approximations which contributes to better understanding of political orientation. 
    %We can extend this study by including methods for learning graph representations for multimodal bias detection.
    \item \textbf{Direction - 3:} Knowledge infusion~\cite{farimani2021leveraging,10.1145/3543507.3583300} and few-shot learning methods have a promising future direction in characterizing political orientation of the given text data. This is specifically true given that the distribution of labels is unpredictable for each social media forum and it is difficult to obtain labeled social media posts. 
\end{enumerate}
%In the future, we can extend this study by including methods for learning graph representations for multimodal bias detection.

\section{Declarations}

\subsection{Ethical Approval}
Not Applicable
\subsection{Availability of supporting data}
We publish all supporting data used in the experiments of this paper at a github page\footnote{\url{https://github.com/sadiakamal/Tweets_Political_Orientation/tree/main}}.
\subsection{Competing Interests}
No, the authors do not have any competing interests as defined by Springer.
\subsection{Funding}
This paper is partly funded by the NSF Research Experience for Undergraduates (REU) grant (\emph{Award No.: 2050978})

\subsection{Author's Contributions}
All authors of this paper are involved in the following categories of writing, data preprocessing, and running experiments.

\begin{enumerate}
\item \emph{Sadia Kamal} is a Ph.D. student at Oklahoma State University. She contributes in directing undergraduate students involved in this research, evaluating baseline results given by undergraduate students, developing label heuristic approaches, running experiments with BERT and label evaluation, and writing Related work, Datasets, Methodology, and Results sections

\item \emph{Brenner Little} is an undergraduate student at Oklahoma State University. He contributes in preparing datasets and running experiments with few-shot learning. Brenner Little is funded by NSF REU program to participate in this research

\item \emph{Jade Gullic} is an undergraduate student at Oklahoma State University. She contributes in data preprocessing and cleaning the data for all experiments, and developing programs to label the data with the proposed heuristic approaches. Jade Gullic is funded by NSF REU program to participate in this research

\item \emph{Trevor Harms} is an undergradaute student at Southwestern Oklahoma State University. He contributes in running all baseline experiments with CountVectorizer, TFIDFVectorizer, and Word2Vec together with classifier models. Trevor Harms is funded by NSF REU program to participate in this research

\item \emph{Kristin Olofsson} is an Assistant Professor at Colorado State University. She contributes to this paper in developing hand-curated dataset for label evaluation experiments

\item \emph{Arunkumar Bagavathi} is an Assistant Professor at Oklahoma State University. He contributes to this paper by collecting the raw dataset required for students, writing Abstract, Introduction, amd Conclusion sections of the paper, and guiding all students in the research

\end{enumerate}

\subsection{Acknowledgements}
Not Applicable

%
%
%

%
% ---- Bibliography ----
%
% BibTeX users should specify bibliography style 'splncs04'.
% References will then be sorted and formatted in the correct style.
%
%\bibliographystyle{splncs04}
\bibliography{bibliography}

%% BioMed_Central_Bib_Style_v1.01

\begin{thebibliography}{43}
% BibTex style file: bmc-mathphys.bst (version 2.1), 2014-07-24
\ifx \bisbn   \undefined \def \bisbn  #1{ISBN #1}\fi
\ifx \binits  \undefined \def \binits#1{#1}\fi
\ifx \bauthor  \undefined \def \bauthor#1{#1}\fi
\ifx \batitle  \undefined \def \batitle#1{#1}\fi
\ifx \bjtitle  \undefined \def \bjtitle#1{#1}\fi
\ifx \bvolume  \undefined \def \bvolume#1{\textbf{#1}}\fi
\ifx \byear  \undefined \def \byear#1{#1}\fi
\ifx \bissue  \undefined \def \bissue#1{#1}\fi
\ifx \bfpage  \undefined \def \bfpage#1{#1}\fi
\ifx \blpage  \undefined \def \blpage #1{#1}\fi
\ifx \burl  \undefined \def \burl#1{\textsf{#1}}\fi
\ifx \doiurl  \undefined \def \doiurl#1{\url{https://doi.org/#1}}\fi
\ifx \betal  \undefined \def \betal{\textit{et al.}}\fi
\ifx \binstitute  \undefined \def \binstitute#1{#1}\fi
\ifx \binstitutionaled  \undefined \def \binstitutionaled#1{#1}\fi
\ifx \bctitle  \undefined \def \bctitle#1{#1}\fi
\ifx \beditor  \undefined \def \beditor#1{#1}\fi
\ifx \bpublisher  \undefined \def \bpublisher#1{#1}\fi
\ifx \bbtitle  \undefined \def \bbtitle#1{#1}\fi
\ifx \bedition  \undefined \def \bedition#1{#1}\fi
\ifx \bseriesno  \undefined \def \bseriesno#1{#1}\fi
\ifx \blocation  \undefined \def \blocation#1{#1}\fi
\ifx \bsertitle  \undefined \def \bsertitle#1{#1}\fi
\ifx \bsnm \undefined \def \bsnm#1{#1}\fi
\ifx \bsuffix \undefined \def \bsuffix#1{#1}\fi
\ifx \bparticle \undefined \def \bparticle#1{#1}\fi
\ifx \barticle \undefined \def \barticle#1{#1}\fi
\bibcommenthead
\ifx \bconfdate \undefined \def \bconfdate #1{#1}\fi
\ifx \botherref \undefined \def \botherref #1{#1}\fi
\ifx \url \undefined \def \url#1{\textsf{#1}}\fi
\ifx \bchapter \undefined \def \bchapter#1{#1}\fi
\ifx \bbook \undefined \def \bbook#1{#1}\fi
\ifx \bcomment \undefined \def \bcomment#1{#1}\fi
\ifx \oauthor \undefined \def \oauthor#1{#1}\fi
\ifx \citeauthoryear \undefined \def \citeauthoryear#1{#1}\fi
\ifx \endbibitem  \undefined \def \endbibitem {}\fi
\ifx \bconflocation  \undefined \def \bconflocation#1{#1}\fi
\ifx \arxivurl  \undefined \def \arxivurl#1{\textsf{#1}}\fi
\csname PreBibitemsHook\endcsname

%%% 1
\bibitem[\protect\citeauthoryear{Sikder et~al.}{2020}]{sikder2020minimalistic}
\begin{barticle}
\bauthor{\bsnm{Sikder}, \binits{O.}},
\bauthor{\bsnm{Smith}, \binits{R.E.}},
\bauthor{\bsnm{Vivo}, \binits{P.}},
\bauthor{\bsnm{Livan}, \binits{G.}}:
\batitle{A minimalistic model of bias, polarization and misinformation in social networks}.
\bjtitle{Scientific reports}
\bvolume{10}(\bissue{1}),
\bfpage{1}--\blpage{11}
(\byear{2020})
\end{barticle}
\endbibitem

%%% 2
\bibitem[\protect\citeauthoryear{Pierri et~al.}{2023}]{10.1145/3578503.3583597}
\begin{bchapter}
\bauthor{\bsnm{Pierri}, \binits{F.}},
\bauthor{\bsnm{Luceri}, \binits{L.}},
\bauthor{\bsnm{Jindal}, \binits{N.}},
\bauthor{\bsnm{Ferrara}, \binits{E.}}:
\bctitle{Propaganda and misinformation on facebook and twitter during the russian invasion of ukraine}.
\bsertitle{WebSci '23},
pp. \bfpage{65}--\blpage{74}.
\bpublisher{Association for Computing Machinery},
\blocation{New York, NY, USA}
(\byear{2023})
\end{bchapter}
\endbibitem

%%% 3
\bibitem[\protect\citeauthoryear{Bernhardt et~al.}{2008}]{bernhardt2008political}
\begin{barticle}
\bauthor{\bsnm{Bernhardt}, \binits{D.}},
\bauthor{\bsnm{Krasa}, \binits{S.}},
\bauthor{\bsnm{Polborn}, \binits{M.}}:
\batitle{Political polarization and the electoral effects of media bias}.
\bjtitle{Journal of Public Economics}
\bvolume{92}(\bissue{5-6}),
\bfpage{1092}--\blpage{1104}
(\byear{2008})
\end{barticle}
\endbibitem

%%% 4
\bibitem[\protect\citeauthoryear{Nair et~al.}{2021}]{nair2021diffusion}
\begin{barticle}
\bauthor{\bsnm{Nair}, \binits{S.}},
\bauthor{\bsnm{Ng}, \binits{K.W.}},
\bauthor{\bsnm{Iamnitchi}, \binits{A.}},
\bauthor{\bsnm{Skvoretz}, \binits{J.}}:
\batitle{Diffusion of social conventions across polarized communities: an empirical study}.
\bjtitle{SNAM}
\bvolume{11}(\bissue{1}),
\bfpage{1}--\blpage{17}
(\byear{2021})
\end{barticle}
\endbibitem

%%% 5
\bibitem[\protect\citeauthoryear{Vicario et~al.}{2019}]{vicario2019polarization}
\begin{barticle}
\bauthor{\bsnm{Vicario}, \binits{M.D.}},
\bauthor{\bsnm{Quattrociocchi}, \binits{W.}},
\bauthor{\bsnm{Scala}, \binits{A.}},
\bauthor{\bsnm{Zollo}, \binits{F.}}:
\batitle{Polarization and fake news: Early warning of potential misinformation targets}.
\bjtitle{ACM TWEB}
\bvolume{13}(\bissue{2}),
\bfpage{1}--\blpage{22}
(\byear{2019})
\end{barticle}
\endbibitem

%%% 6
\bibitem[\protect\citeauthoryear{Fagni and Cresci}{2022}]{fagni2022fine}
\begin{barticle}
\bauthor{\bsnm{Fagni}, \binits{T.}},
\bauthor{\bsnm{Cresci}, \binits{S.}}:
\batitle{Fine-grained prediction of political leaning on social media with unsupervised deep learning}.
\bjtitle{Journal of Artificial Intelligence Research}
\bvolume{73},
\bfpage{633}--\blpage{672}
(\byear{2022})
\end{barticle}
\endbibitem

%%% 7
\bibitem[\protect\citeauthoryear{A{\"\i}meur et~al.}{2023}]{aimeur2023fake}
\begin{barticle}
\bauthor{\bsnm{A{\"\i}meur}, \binits{E.}},
\bauthor{\bsnm{Amri}, \binits{S.}},
\bauthor{\bsnm{Brassard}, \binits{G.}}:
\batitle{Fake news, disinformation and misinformation in social media: a review}.
\bjtitle{Social Network Analysis and Mining}
\bvolume{13}(\bissue{1}),
\bfpage{30}
(\byear{2023})
\end{barticle}
\endbibitem

%%% 8
\bibitem[\protect\citeauthoryear{Bagavathi et~al.}{2019}]{bagavathi2019examining}
\begin{bchapter}
\bauthor{\bsnm{Bagavathi}, \binits{A.}},
\bauthor{\bsnm{Bashiri}, \binits{P.}},
\bauthor{\bsnm{Reid}, \binits{S.}},
\bauthor{\bsnm{Phillips}, \binits{M.}},
\bauthor{\bsnm{Krishnan}, \binits{S.}}:
\bctitle{Examining untempered social media: analyzing cascades of polarized conversations}.
In: \bbtitle{IEEE/ACM ASONAM},
pp. \bfpage{625}--\blpage{632}
(\byear{2019})
\end{bchapter}
\endbibitem

%%% 9
\bibitem[\protect\citeauthoryear{Brena et~al.}{2019}]{brena2019news}
\begin{bchapter}
\bauthor{\bsnm{Brena}, \binits{G.}},
\bauthor{\bsnm{Brambilla}, \binits{M.}},
\bauthor{\bsnm{Ceri}, \binits{S.}},
\bauthor{\bsnm{Di~Giovanni}, \binits{M.}},
\bauthor{\bsnm{Pierri}, \binits{F.}},
\bauthor{\bsnm{Ramponi}, \binits{G.}}:
\bctitle{News sharing user behaviour on twitter: a comprehensive data collection of news articles and social interactions}.
In: \bbtitle{Proceedings of the International AAAI Conference on Web and Social Media},
vol. \bseriesno{13},
pp. \bfpage{592}--\blpage{597}
(\byear{2019})
\end{bchapter}
\endbibitem

%%% 10
\bibitem[\protect\citeauthoryear{Fair and Wesslen}{2019}]{fair2019shouting}
\begin{bchapter}
\bauthor{\bsnm{Fair}, \binits{G.}},
\bauthor{\bsnm{Wesslen}, \binits{R.}}:
\bctitle{Shouting into the void: A database of the alternative social media platform gab}.
In: \bbtitle{Proceedings of the International AAAI Conference on Web and Social Media},
vol. \bseriesno{13},
pp. \bfpage{608}--\blpage{610}
(\byear{2019})
\end{bchapter}
\endbibitem

%%% 11
\bibitem[\protect\citeauthoryear{Belcastro et~al.}{2020}]{9026882}
\begin{barticle}
\bauthor{\bsnm{Belcastro}, \binits{L.}},
\bauthor{\bsnm{Cantini}, \binits{R.}},
\bauthor{\bsnm{Marozzo}, \binits{F.}},
\bauthor{\bsnm{Talia}, \binits{D.}},
\bauthor{\bsnm{Trunfio}, \binits{P.}}:
\batitle{Learning political polarization on social media using neural networks}.
\bjtitle{IEEE Access}
\bvolume{8},
\bfpage{47177}--\blpage{47187}
(\byear{2020})
\end{barticle}
\endbibitem

%%% 12
\bibitem[\protect\citeauthoryear{Garimella et~al.}{2021}]{garimella2021political}
\begin{bchapter}
\bauthor{\bsnm{Garimella}, \binits{K.}},
\bauthor{\bsnm{Smith}, \binits{T.}},
\bauthor{\bsnm{Weiss}, \binits{R.}},
\bauthor{\bsnm{West}, \binits{R.}}:
\bctitle{Political polarization in online news consumption}.
In: \bbtitle{Proceedings of the International AAAI Conference on Web and Social Media},
vol. \bseriesno{15},
pp. \bfpage{152}--\blpage{162}
(\byear{2021})
\end{bchapter}
\endbibitem

%%% 13
\bibitem[\protect\citeauthoryear{Lei et~al.}{2022}]{lei2022sentence}
\begin{bchapter}
\bauthor{\bsnm{Lei}, \binits{Y.}},
\bauthor{\bsnm{Huang}, \binits{R.}},
\bauthor{\bsnm{Wang}, \binits{L.}},
\bauthor{\bsnm{Beauchamp}, \binits{N.}}:
\bctitle{Sentence-level media bias analysis informed by discourse structures}.
In: \bbtitle{Proceedings of the 2022 Conference on Empirical Methods in Natural Language Processing},
pp. \bfpage{10040}--\blpage{10050}
(\byear{2022})
\end{bchapter}
\endbibitem

%%% 14
\bibitem[\protect\citeauthoryear{Sinno et~al.}{2022}]{sinno2022political}
\begin{bchapter}
\bauthor{\bsnm{Sinno}, \binits{B.}},
\bauthor{\bsnm{Oviedo}, \binits{B.}},
\bauthor{\bsnm{Atwell}, \binits{K.}},
\bauthor{\bsnm{Alikhani}, \binits{M.}},
\bauthor{\bsnm{Li}, \binits{J.J.}}:
\bctitle{Political ideology and polarization: A multi-dimensional approach}.
In: \bbtitle{Proceedings of the 2022 Conference of the North American Chapter of the Association for Computational Linguistics: Human Language Technologies},
pp. \bfpage{231}--\blpage{243}
(\byear{2022})
\end{bchapter}
\endbibitem

%%% 15
\bibitem[\protect\citeauthoryear{Conover et~al.}{}]{6113114}
\begin{botherref}
\oauthor{\bsnm{Conover}, \binits{M.D.}},
\oauthor{\bsnm{Goncalves}, \binits{B.}},
\oauthor{\bsnm{Ratkiewicz}, \binits{J.}},
\oauthor{\bsnm{Flammini}, \binits{A.}},
\oauthor{\bsnm{Menczer}, \binits{F.}}:
Predicting the political alignment of twitter users.
In: 2011 IEEE Third International Conference on Privacy, Security, Risk and Trust and 2011 IEEE Third International Conference on Social Computing,
pp. 192--199
\end{botherref}
\endbibitem

%%% 16
\bibitem[\protect\citeauthoryear{Chen et~al.}{2017}]{10.5555/3172077.3172399}
\begin{bchapter}
\bauthor{\bsnm{Chen}, \binits{W.}},
\bauthor{\bsnm{Zhang}, \binits{X.}},
\bauthor{\bsnm{Wang}, \binits{T.}},
\bauthor{\bsnm{Yang}, \binits{B.}},
\bauthor{\bsnm{Li}, \binits{Y.}}:
\bctitle{Opinion-aware knowledge graph for political ideology detection}.
In: \bbtitle{Proceedings of the 26th International Joint Conference on Artificial Intelligence}.
\bsertitle{IJCAI'17},
pp. \bfpage{3647}--\blpage{3653}.
\bpublisher{AAAI Press}, \blocation{???}
(\byear{2017})
\end{bchapter}
\endbibitem

%%% 17
\bibitem[\protect\citeauthoryear{Lyu and Luo}{2022}]{10.1145/3503161.3547898}
\begin{bchapter}
\bauthor{\bsnm{Lyu}, \binits{H.}},
\bauthor{\bsnm{Luo}, \binits{J.}}:
\bctitle{Understanding political polarization via jointly modeling users, connections and multimodal contents on heterogeneous graphs}.
In: \bbtitle{Proceedings of the 30th ACM International Conference on Multimedia}.
\bsertitle{MM '22},
pp. \bfpage{4072}--\blpage{4082}.
\bpublisher{Association for Computing Machinery},
\blocation{New York, NY, USA}
(\byear{2022})
\end{bchapter}
\endbibitem

%%% 18
\bibitem[\protect\citeauthoryear{Preo{\c{t}}iuc-Pietro et~al.}{2017}]{preotiuc-pietro-etal-2017-beyond}
\begin{bchapter}
\bauthor{\bsnm{Preo{\c{t}}iuc-Pietro}, \binits{D.}},
\bauthor{\bsnm{Liu}, \binits{Y.}},
\bauthor{\bsnm{Hopkins}, \binits{D.}},
\bauthor{\bsnm{Ungar}, \binits{L.}}:
\bctitle{Beyond binary labels: Political ideology prediction of {T}witter users}.
In: \bbtitle{Proceedings of the 55th Annual Meeting of the Association for Computational Linguistics (Volume 1: Long Papers)},
\bconflocation{Vancouver, Canada},
pp. \bfpage{729}--\blpage{740}
(\byear{2017})
\end{bchapter}
\endbibitem

%%% 19
\bibitem[\protect\citeauthoryear{Jiang et~al.}{2023}]{jiang2023retweet}
\begin{bchapter}
\bauthor{\bsnm{Jiang}, \binits{J.}},
\bauthor{\bsnm{Ren}, \binits{X.}},
\bauthor{\bsnm{Ferrara}, \binits{E.}}:
\bctitle{Retweet-bert: political leaning detection using language features and information diffusion on social networks}.
In: \bbtitle{Proceedings of the International AAAI Conference on Web and Social Media},
vol. \bseriesno{17},
pp. \bfpage{459}--\blpage{469}
(\byear{2023})
\end{bchapter}
\endbibitem

%%% 20
\bibitem[\protect\citeauthoryear{Hosseinmardi et~al.}{2021}]{hosseinmardi2021examining}
\begin{botherref}
\oauthor{\bsnm{Hosseinmardi}, \binits{H.}},
\oauthor{\bsnm{Ghasemian}, \binits{A.}},
\oauthor{\bsnm{Clauset}, \binits{A.}},
\oauthor{\bsnm{Mobius}, \binits{M.}},
\oauthor{\bsnm{Rothschild}, \binits{D.M.}},
\oauthor{\bsnm{Watts}, \binits{D.J.}}:
Examining the consumption of radical content on youtube.
Proceedings of the National Academy of Sciences
\textbf{118}(32)
(2021)
\end{botherref}
\endbibitem

%%% 21
\bibitem[\protect\citeauthoryear{Gerrish and Blei}{2011}]{10.5555/3104482.3104544}
\begin{bchapter}
\bauthor{\bsnm{Gerrish}, \binits{S.M.}},
\bauthor{\bsnm{Blei}, \binits{D.M.}}:
\bctitle{Predicting legislative roll calls from text}.
In: \bbtitle{International Conference on Machine Learning(ICML)},
pp. \bfpage{489}--\blpage{496}
(\byear{2011})
\end{bchapter}
\endbibitem

%%% 22
\bibitem[\protect\citeauthoryear{Olorunnimbe and Viktor}{2015}]{olorunnimbe2015tweets}
\begin{bchapter}
\bauthor{\bsnm{Olorunnimbe}, \binits{M.K.}},
\bauthor{\bsnm{Viktor}, \binits{H.L.}}:
\bctitle{Tweets as a vote: Exploring political sentiments on twitter for opinion mining}.
In: \bbtitle{Foundations of Intelligent Systems: 22nd International Symposium, ISMIS 2015, Lyon, France, October 21-23, 2015, Proceedings 22},
pp. \bfpage{180}--\blpage{185}
(\byear{2015}).
\bcomment{Springer}
\end{bchapter}
\endbibitem

%%% 23
\bibitem[\protect\citeauthoryear{Chen et~al.}{2018}]{chen-etal-2018-learning}
\begin{bchapter}
\bauthor{\bsnm{Chen}, \binits{W.-F.}},
\bauthor{\bsnm{Wachsmuth}, \binits{H.}},
\bauthor{\bsnm{Al-Khatib}, \binits{K.}},
\bauthor{\bsnm{Stein}, \binits{B.}}:
\bctitle{Learning to flip the bias of news headlines}.
In: \bbtitle{11th ACL International Conference on Natural Language Generation},
pp. \bfpage{79}--\blpage{88}
(\byear{2018})
\end{bchapter}
\endbibitem

%%% 24
\bibitem[\protect\citeauthoryear{Kulkarni et~al.}{2018}]{kulkarni-etal-2018-multi}
\begin{bchapter}
\bauthor{\bsnm{Kulkarni}, \binits{V.}},
\bauthor{\bsnm{Ye}, \binits{J.}},
\bauthor{\bsnm{Skiena}, \binits{S.}},
\bauthor{\bsnm{Wang}, \binits{W.Y.}}:
\bctitle{Multi-view models for political ideology detection of news articles}.
In: \bbtitle{Proceedings of the 2018 Conference on Empirical Methods in Natural Language Processing},
pp. \bfpage{3518}--\blpage{3527}.
\bpublisher{Association for Computational Linguistics},
\blocation{Brussels, Belgium}
(\byear{2018})
\end{bchapter}
\endbibitem

%%% 25
\bibitem[\protect\citeauthoryear{Xiao et~al.}{2023}]{xiao2023detecting}
\begin{barticle}
\bauthor{\bsnm{Xiao}, \binits{Z.}},
\bauthor{\bsnm{Zhu}, \binits{J.}},
\bauthor{\bsnm{Wang}, \binits{Y.}},
\bauthor{\bsnm{Zhou}, \binits{P.}},
\bauthor{\bsnm{Lam}, \binits{W.H.}},
\bauthor{\bsnm{Porter}, \binits{M.A.}},
\bauthor{\bsnm{Sun}, \binits{Y.}}:
\batitle{Detecting political biases of named entities and hashtags on twitter}.
\bjtitle{EPJ Data Science}
\bvolume{12}(\bissue{1}),
\bfpage{20}
(\byear{2023})
\end{barticle}
\endbibitem

%%% 26
\bibitem[\protect\citeauthoryear{Pandya et~al.}{2022}]{pandya2022proposal}
\begin{bchapter}
\bauthor{\bsnm{Pandya}, \binits{K.J.}},
\bauthor{\bsnm{Jaiswal}, \binits{A.}},
\bauthor{\bsnm{Rautaray}, \binits{S.S.}},
\bauthor{\bsnm{Pandey}, \binits{M.}}:
\bctitle{A proposal to find fake news and detecting political bias of news articles}.
In: \bbtitle{Advances in Data and Information Sciences},
pp. \bfpage{515}--\blpage{526}
(\byear{2022}).
\bcomment{Springer}
\end{bchapter}
\endbibitem

%%% 27
\bibitem[\protect\citeauthoryear{Glazkova}{2021}]{glazkova2021comparison}
\begin{botherref}
\oauthor{\bsnm{Glazkova}, \binits{A.}}:
A comparison of text representation methods for predicting political views of social media users.
Proc. of Information Technologies and Intelligent Decision Making Systems
(2021)
\end{botherref}
\endbibitem

%%% 28
\bibitem[\protect\citeauthoryear{Cruz et~al.}{2020}]{cruz2020document}
\begin{bchapter}
\bauthor{\bsnm{Cruz}, \binits{A.F.}},
\bauthor{\bsnm{Rocha}, \binits{G.}},
\bauthor{\bsnm{Cardoso}, \binits{H.L.}}:
\bctitle{On document representations for detection of biased news articles}.
In: \bbtitle{Proceedings of the 35th Annual ACM Symposium on Applied Computing},
pp. \bfpage{892}--\blpage{899}
(\byear{2020})
\end{bchapter}
\endbibitem

%%% 29
\bibitem[\protect\citeauthoryear{Xiao et~al.}{2020}]{osti_10178656}
\begin{botherref}
\oauthor{\bsnm{Xiao}, \binits{Z.}},
\oauthor{\bsnm{Song}, \binits{W.}},
\oauthor{\bsnm{Xu}, \binits{H.}},
\oauthor{\bsnm{Ren}, \binits{Z.}},
\oauthor{\bsnm{Sun}, \binits{Y.}}:
Timme: Twitter ideology-detection via multi-task multi-relational embedding.
ACM SIGKDD,
2258--2268
(2020)
\end{botherref}
\endbibitem

%%% 30
\bibitem[\protect\citeauthoryear{Sales et~al.}{2019}]{sales2019media}
\begin{bchapter}
\bauthor{\bsnm{Sales}, \binits{A.}},
\bauthor{\bsnm{Balby}, \binits{L.}},
\bauthor{\bsnm{Veloso}, \binits{A.}}:
\bctitle{Media bias characterization in brazilian presidential elections}.
In: \bbtitle{ACM Conference on Hypertext and Social Media},
pp. \bfpage{231}--\blpage{240}
(\byear{2019})
\end{bchapter}
\endbibitem

%%% 31
\bibitem[\protect\citeauthoryear{Baly et~al.}{2020}]{baly-etal-2020-detect}
\begin{bchapter}
\bauthor{\bsnm{Baly}, \binits{R.}},
\bauthor{\bsnm{Da~San~Martino}, \binits{G.}},
\bauthor{\bsnm{Glass}, \binits{J.}},
\bauthor{\bsnm{Nakov}, \binits{P.}}:
\bctitle{We can detect your bias: Predicting the political ideology of news articles}.
In: \bbtitle{Proceedings of the 2020 Conference on Empirical Methods in Natural Language Processing (EMNLP)},
pp. \bfpage{4982}--\blpage{4991}.
\bpublisher{Association for Computational Linguistics},
\blocation{Online}
(\byear{2020})
\end{bchapter}
\endbibitem

%%% 32
\bibitem[\protect\citeauthoryear{Baly et~al.}{2019}]{baly-etal-2019-multi}
\begin{bchapter}
\bauthor{\bsnm{Baly}, \binits{R.}},
\bauthor{\bsnm{Karadzhov}, \binits{G.}},
\bauthor{\bsnm{Saleh}, \binits{A.}},
\bauthor{\bsnm{Glass}, \binits{J.}},
\bauthor{\bsnm{Nakov}, \binits{P.}}:
\bctitle{Multi-task ordinal regression for jointly predicting the trustworthiness and the leading political ideology of news media}.
In: \bbtitle{Proceedings of the 2019 Conference of the North {A}merican Chapter of the Association for Computational Linguistics: Human Language Technologies, Volume 1 (Long and Short Papers)},
pp. \bfpage{2109}--\blpage{2116}.
\bpublisher{Association for Computational Linguistics},
\blocation{Minneapolis, Minnesota}
(\byear{2019})
\end{bchapter}
\endbibitem

%%% 33
\bibitem[\protect\citeauthoryear{Gangula et~al.}{2019}]{gangula-etal-2019-detecting}
\begin{bchapter}
\bauthor{\bsnm{Gangula}, \binits{R.R.R.}},
\bauthor{\bsnm{Duggenpudi}, \binits{S.R.}},
\bauthor{\bsnm{Mamidi}, \binits{R.}}:
\bctitle{Detecting political bias in news articles using headline attention}.
In: \bbtitle{ACL Workshop BlackboxNLP: Analyzing and Interpreting Neural Networks for NLP},
pp. \bfpage{77}--\blpage{84}
(\byear{2019})
\end{bchapter}
\endbibitem

%%% 34
\bibitem[\protect\citeauthoryear{Ren et~al.}{2022}]{ren2022discrimination}
\begin{bchapter}
\bauthor{\bsnm{Ren}, \binits{Y.}},
\bauthor{\bsnm{Liu}, \binits{Y.}},
\bauthor{\bsnm{Zhang}, \binits{G.}},
\bauthor{\bsnm{Liu}, \binits{L.}},
\bauthor{\bsnm{Lv}, \binits{P.}}:
\bctitle{Discrimination of news political bias based on heterogeneous graph neural network}.
In: \bbtitle{Knowledge Science, Engineering and Management: 15th International Conference, KSEM 2022, Singapore, August 6--8, 2022, Proceedings, Part I},
pp. \bfpage{542}--\blpage{555}
(\byear{2022}).
\bcomment{Springer}
\end{bchapter}
\endbibitem

%%% 35
\bibitem[\protect\citeauthoryear{Ko et~al.}{2023}]{10.1145/3543507.3583300}
\begin{bchapter}
\bauthor{\bsnm{Ko}, \binits{Y.}},
\bauthor{\bsnm{Ryu}, \binits{S.}},
\bauthor{\bsnm{Han}, \binits{S.}},
\bauthor{\bsnm{Jeon}, \binits{Y.}},
\bauthor{\bsnm{Kim}, \binits{J.}},
\bauthor{\bsnm{Park}, \binits{S.}},
\bauthor{\bsnm{Han}, \binits{K.}},
\bauthor{\bsnm{Tong}, \binits{H.}},
\bauthor{\bsnm{Kim}, \binits{S.-W.}}:
\bctitle{Khan: Knowledge-aware hierarchical attention networks for accurate political stance prediction}.
In: \bbtitle{Proceedings of the ACM Web Conference 2023}.
\bsertitle{WWW '23},
pp. \bfpage{1572}--\blpage{1583}.
\bpublisher{Association for Computing Machinery},
\blocation{New York, NY, USA}
(\byear{2023})
\end{bchapter}
\endbibitem

%%% 36
\bibitem[\protect\citeauthoryear{Chakraborty et~al.}{2022}]{chakraborty2022fast}
\begin{bchapter}
\bauthor{\bsnm{Chakraborty}, \binits{S.}},
\bauthor{\bsnm{Goyal}, \binits{P.}},
\bauthor{\bsnm{Mukherjee}, \binits{A.}}:
\bctitle{Fast few shot self-attentive semi-supervised political inclination prediction}.
In: \bbtitle{24th International Conference on Asian Digital Libraries, ICADL 2022, Hanoi, Vietnam, November 30--December 2, 2022, Proceedings},
pp. \bfpage{3}--\blpage{20}
(\byear{2022}).
\bcomment{Springer}
\end{bchapter}
\endbibitem

%%% 37
\bibitem[\protect\citeauthoryear{Snell et~al.}{2017}]{snell2017prototypical}
\begin{botherref}
\oauthor{\bsnm{Snell}, \binits{J.}},
\oauthor{\bsnm{Swersky}, \binits{K.}},
\oauthor{\bsnm{Zemel}, \binits{R.}}:
Prototypical networks for few-shot learning.
In: Advances in neural information processing systems
\textbf{30}
(2017)
\end{botherref}
\endbibitem

%%% 38
\bibitem[\protect\citeauthoryear{Shahmirzadi et~al.}{2019}]{shahmirzadi2019text}
\begin{bchapter}
\bauthor{\bsnm{Shahmirzadi}, \binits{O.}},
\bauthor{\bsnm{Lugowski}, \binits{A.}},
\bauthor{\bsnm{Younge}, \binits{K.}}:
\bctitle{Text similarity in vector space models: a comparative study}.
In: \bbtitle{IEEE International Conference on Machine Learning and Applications (ICMLA)},
pp. \bfpage{659}--\blpage{666}
(\byear{2019})
\end{bchapter}
\endbibitem

%%% 39
\bibitem[\protect\citeauthoryear{Ramos et~al.}{2003}]{ramos2003using}
\begin{bchapter}
\bauthor{\bsnm{Ramos}, \binits{J.}}, \betal:
\bctitle{Using tf-idf to determine word relevance in document queries}.
In: \bbtitle{Proceedings of the First Instructional Conference on Machine Learning},
vol. \bseriesno{242},
pp. \bfpage{29}--\blpage{48}
(\byear{2003})
\end{bchapter}
\endbibitem

%%% 40
\bibitem[\protect\citeauthoryear{Mikolov et~al.}{2013}]{DBLP:journals/corr/abs-1301-3781}
\begin{bchapter}
\bauthor{\bsnm{Mikolov}, \binits{T.}},
\bauthor{\bsnm{Chen}, \binits{K.}},
\bauthor{\bsnm{Corrado}, \binits{G.}},
\bauthor{\bsnm{Dean}, \binits{J.}}:
\bctitle{Efficient estimation of word representations in vector space}.
In: \bbtitle{Workshop Proceedings of International Conference on Learning Representations (ICLR)}
(\byear{2013})
\end{bchapter}
\endbibitem

%%% 41
\bibitem[\protect\citeauthoryear{Devlin et~al.}{2019}]{devlin-etal-2019-bert}
\begin{bchapter}
\bauthor{\bsnm{Devlin}, \binits{J.}},
\bauthor{\bsnm{Chang}, \binits{M.-W.}},
\bauthor{\bsnm{Lee}, \binits{K.}},
\bauthor{\bsnm{Toutanova}, \binits{K.}}:
\bctitle{{BERT}: Pre-training of deep bidirectional transformers for language understanding}.
In: \bbtitle{Proceedings of the 2019 Conference of the North {A}merican Chapter of the Association for Computational Linguistics: Human Language Technologies},
pp. \bfpage{4171}--\blpage{4186}.
\bpublisher{Association for Computational Linguistics},
\blocation{Minneapolis, Minnesota}
(\byear{2019})
\end{bchapter}
\endbibitem

%%% 42
\bibitem[\protect\citeauthoryear{Lewis et~al.}{2020}]{lewis-etal-2020-bart}
\begin{bchapter}
\bauthor{\bsnm{Lewis}, \binits{M.}},
\bauthor{\bsnm{Liu}, \binits{Y.}},
\bauthor{\bsnm{Goyal}, \binits{N.}},
\bauthor{\bsnm{Ghazvininejad}, \binits{M.}},
\bauthor{\bsnm{Mohamed}, \binits{A.}},
\bauthor{\bsnm{Levy}, \binits{O.}},
\bauthor{\bsnm{Stoyanov}, \binits{V.}},
\bauthor{\bsnm{Zettlemoyer}, \binits{L.}}:
\bctitle{{BART}: Denoising sequence-to-sequence pre-training for natural language generation, translation, and comprehension}.
In: \bbtitle{Proceedings of the 58th Annual Meeting of the Association for Computational Linguistics},
pp. \bfpage{7871}--\blpage{7880}
(\byear{2020})
\end{bchapter}
\endbibitem

%%% 43
\bibitem[\protect\citeauthoryear{Farimani et~al.}{2021}]{farimani2021leveraging}
\begin{bchapter}
\bauthor{\bsnm{Farimani}, \binits{S.A.}},
\bauthor{\bsnm{Jahan}, \binits{M.V.}},
\bauthor{\bsnm{Fard}, \binits{A.M.}},
\bauthor{\bsnm{Haffari}, \binits{G.}}:
\bctitle{Leveraging latent economic concepts and sentiments in the news for market prediction}.
In: \bbtitle{IEEE DSAA},
pp. \bfpage{1}--\blpage{10}
(\byear{2021})
\end{bchapter}
\endbibitem

\end{thebibliography}

\end{document}